\documentclass[tikz,11pt,english,a4paper]{article}
\usepackage{float}
\usepackage{jcappub}
\usepackage[utf8]{inputenc}

\usepackage{bm}
\usepackage{algorithm}
\usepackage{nicefrac}
\usepackage{bbold}
\usepackage{dsfont}
\usepackage{setspace}
\usepackage[linewidth=1pt]{mdframed}
\usepackage{bold-extra}
\usepackage{tabto}

\usepackage[noend]{algpseudocode}
\makeatletter
\def\BState{\State\hskip-\ALG@thistlm}
\makeatother

\usepackage{fontawesome5}
\usepackage{hyperref}
\makeatletter
\newcommand{\github}[1]{%
   \href{#1}{\faGithub}%
}
\makeatother

\newcommand{\connect}{\textsc{connect}}
\newcommand{\class}{\textsc{class}}
\newcommand{\camb}{\textsc{camb}}
\newcommand{\montepython}{\textsc{MontePython}}

\usepackage{siunitx}
\DeclareSIUnit \parsec {pc}

\usepackage[T1]{fontenc}

\usepackage{lmodern} % To switch to Latin Modern
\rmfamily % To load Latin Modern Roman and enable the following NFSS declarations.
\DeclareFontShape{T1}{lmr}{b}{sc}{<->ssub*cmr/bx/sc}{}
\DeclareFontShape{T1}{lmr}{bx}{sc}{<->ssub*cmr/bx/sc}{}

\usepackage{soul}

\usepackage[edges]{forest}
\definecolor{folderbg}{RGB}{124,166,198}
\definecolor{folderborder}{RGB}{110,144,169}
\newlength\Size
\tikzset{%
  folder/.pic={%
    \filldraw [draw=folderborder, top color=folderbg!50, bottom color=folderbg] (-1.05*\Size,0.2\Size+5pt) rectangle ++(.75*\Size,-0.2\Size-5pt);
    \filldraw [draw=folderborder, top color=folderbg!50, bottom color=folderbg] (-1.15*\Size,-\Size) rectangle (1.15*\Size,\Size);
  },
  file/.pic={%
    \filldraw [draw=folderborder, top color=folderbg!5, bottom color=folderbg!10] (-\Size,.4*\Size+5pt) coordinate (a) |- (\Size,-1.2*\Size) coordinate (b) -- ++(0,1.6*\Size) coordinate (c) -- ++(-5pt,5pt) coordinate (d) -- cycle (d) |- (c) ;
  },
}
\forestset{%
  declare autowrapped toks={pic me}{},
  declare boolean register={pic root},
  pic root=0,
  pic dir tree/.style={%
    for tree={%
      folder,
      font=\ttfamily,
      grow'=0,
    },
    before typesetting nodes={%
      for tree={%
        edge label+/.option={pic me},
      },
      if pic root={
        tikz+={
          \pic at ([xshift=\Size].west) {folder};
        },
        align={l}
      }{},
    },
  },
  pic me set/.code n args=2{%
    \forestset{%
      #1/.style={%
        inner xsep=2\Size,
        pic me={pic {#2}},
      }
    }
  },
  pic me set={directory}{folder},
  pic me set={file}{file},
}

\usepackage{newfloat}
\usepackage{caption}
\DeclareFloatingEnvironment[fileext=frm,placement={!t},name=Algorithm]{pseudoenv}
\DeclareCaptionLabelSeparator{normal-colon}{\normalfont : } 
\newenvironment{pseudo}[2]{
    \gdef\tempcaption{#1}
    \gdef\templabel{#2}
    \begin{pseudoenv}[tb]
    \begin{mdframed}[roundcorner=10pt, middlelinewidth=1pt]
    \begin{center}
    \begin{tabular}{l|l}}
    {\end{tabular}
    \end{center}
    \end{mdframed}
    \caption{\tempcaption}
    \label{\templabel}
    \end{pseudoenv}
    }

\makeatletter
\g@addto@macro\bfseries{\boldmath}
\makeatother

\makeatletter
\def\old@comma{,}
\catcode`\,=13
\def,{%
  \ifmmode%
    \old@comma\discretionary{}{}{}%
  \else%
    \old@comma%
  \fi%
}
\makeatother

\begin{document}

%%%%%%%%%%%%%%%%%%%%%%%%%%%%%%%%%%%%%%%%%%%%%%%%%%%%%%%%%%%%%%%%%%%%%%
% Frontpage %%%%%%%%%%%%%%%%%%%%%%%%%%%%%%%%%%%%%%%%%%%%%%%%%%%%%%%%%%
%%%%%%%%%%%%%%%%%%%%%%%%%%%%%%%%%%%%%%%%%%%%%%%%%%%%%%%%%%%%%%%%%%%%%%

\title{Cutting corners: Hypersphere sampling as a new standard for cosmological emulators}

\author[a]{Andreas Nygaard,}
\author[a]{Emil Brinch Holm,}
\author[a]{Steen Hannestad,}
\author[a]{and Thomas Tram}

\affiliation[a]{Department of Physics and Astronomy, Aarhus University,
 DK-8000 Aarhus C, Denmark}

\emailAdd{andreas@phys.au.dk}
\emailAdd{ebholm@phys.au.dk}
\emailAdd{steen@phys.au.dk}
\emailAdd{thomas.tram@phys.au.dk}

\abstract{
	Cosmological emulators of observables such as the Cosmic Microwave Background (CMB) spectra and matter power spectra commonly use training data sampled from a Latin hypercube. This method often incurs high computational costs by covering less relevant parts of the parameter space, especially in high dimensions where only a small fraction of the parameter space yields a significant likelihood.
	
	In this paper, we introduce hypersphere sampling, which instead concentrates sample points in regions with higher likelihoods, significantly enhancing the efficiency and accuracy of emulators. A novel algorithm for sampling within a high-dimensional hyperellipsoid aligned with axes of correlation in the cosmological parameters is presented. This method focuses the distribution of training data points on areas of the parameter space that are most relevant to the models being tested, thereby avoiding the computational redundancies common in Latin hypercube approaches.
	
	Comparative analysis using the \connect{} emulation tool demonstrates that hypersphere sampling can achieve similar or improved emulation precision with more than an order of magnitude fewer data points and thus less computational effort than traditional methods. This was tested for both the $\Lambda$CDM model and a 5-parameter extension including Early Dark Energy, massive neutrinos, and additional ultra-relativistic degrees of freedom. Our results suggest that hypersphere sampling holds potential as a more efficient approach for cosmological emulation, particularly suitable for complex, high-dimensional models.
	}

\maketitle
%%%%%%%%%%%%%%%%%%%%%%%%%%%%%%%%%%%%%%%%%%%%%%%%%%%%%%%%%%%%%%%%%%%%%%%%%%%%%%%%%%%%%%%%%%%%%%%%%%

\section{Introduction}\label{sec:introduction}

Using machine learning techniques to emulate observables such as CMB spectra or matter power spectra predicted by a cosmological model has become increasingly popular in recent years, mainly due to the high computational cost of directly computing the observables using standard tools (e.g.\ Einstein-Boltzmann solvers or $N$-body codes). For example, computing Bayesian evidence ratios between different cosmological models typically requires calculating observables in millions of points in the space of cosmological parameters, and using e.g.\ standard codes such as \class{}~\cite{Blas:2011rf} or \camb{}~\cite{Lewis:1999bs} therefore leads to millions of CPU core-seconds being consumed.

This CPU demand can be reduced by orders of magnitude using emulators, often without sacrificing precision. However, there is still a very substantial computational cost related to generating training data, i.e. the predicted observables at each point in the cosmological parameter space, for the emulator. This means that it is important that the training data represent the parameter space well, and necessitates a good balance between the amount of training data points and the relevance of each point. 

A simple choice is to use Latin hypercube sampling of training points on some predefined (prior) volume of parameter space. This has the advantage of being simple to implement and assigning equal weight to all regions within the hypercube.
Examples of this method used in cosmological emulator training include e.g.\ the emulators of Refs.~\cite{SpurioMancini:2021ppk,Bonici:2023xjk,Euclid:2018mlb,Euclid:2020rfv}. 
One could, however, ask why the feature of equal weight to all regions within the prior volume is desired, since most of the volume in higher dimensions is in the corners of the hypercube.  Points in such regions are almost always associated with a very poor likelihood and when using the emulator to calculate for example profile likelihoods, Bayesian parameter inference, or Bayesian evidence, the emulator's ability to accurately calculate observables in these corner regions is wasted. This leads to a very inefficient use of training data and puts substantially higher demands on the number of points in the training data.
Some emulators, such as Refs.~\cite{Gammal:2022eob,Gunther:2023xhh,Ruiz-Zapatero:2023hdf,Angulo:2020vky}, have circumvented this by using Gaussian processes (GP) instead of artificial neural networks. The idea behind this is to let the acquisition function of the GP decide which new point to include in the training data based on where the emulation is uncertain or unexplored. The downside of this is the inferior scalability of GPs compared to neural networks, i.e. only a limited number of dimensions and amount of training data is feasible. Typically, Gaussian processes are therefore used in situations where calculating each individual training point becomes extremely expensive (a good example is emulation of observables based on $N$-body simulations).

In this paper, we illustrate that a better approach when dealing with neural networks is to sample uniformly in a hypersphere, which avoids the large amount of irrelevant points in the corners of the hypercube. This, however, requires prior knowledge about where the region of interest is located in the parameter space, but similar prior knowledge is also needed to construct a Latin hypercube. This kind of hypersphere (or hyperellipsoid) sampling has been used by Refs.~\cite{Gunther:2022pto,Schneider_2011} and discussed in Ref.~\cite{Euclid:2020rfv} but with a rather suboptimal way of sampling by rejecting points from a (Latin) hypercube. In high dimensions this becomes impossible since it requires a Latin hypercube too large to fit in the memory of a computer. There are, however, ways to effectively sample points within a hypersphere, if one omits the "Latin" requirement. 

In section~\ref{sec:sampling}, we explore different sampling strategies and describe a novel, efficient method for sampling uniformly from a high-dimensional hypersphere. Next, we compare performances using the publicly available emulation tool \connect{}~\cite{Nygaard:2022wri} in section~\ref{sec:comparison}, and give our conclusions and outlook in section~\ref{sec:conclusion}.

\section{Sampling methods for training data}\label{sec:sampling}

When building an emulator of Einstein-Boltzmann solvers such as \class{}~\cite{Blas:2011rf} or \camb{}~\cite{Lewis:1999bs}, one generates the training data by running the solver on a selected set of points in parameter space, giving corresponding pairs of cosmological parameters and their corresponding observables at each point. This leaves open the choice of the set of parameter space points at which to generate the data. There are different ways to sample training data depending on what the objective of the network is. Ultimately, a neural network is only as good as its training data, and so a region of sparse data leads to the network having to interpolate over larger distances in this part of the parameter space, while regions of dense data leads to very accurate emulation in these regions. One's choice of training data might depend on various factors such as the number of points one is willing to compute (if each point requires expensive computations such as $N$-body codes, it might not be many) or whether or not the network should be more precise in some regions than others.

In this section, we will go through different ways one might sample data from a hypercube and a hypersphere, and present some of the strengths and drawbacks of each.

\subsection{Latin hypercube sampling}
The idea behind Latin hypercube sampling~\cite{McKay79} is to create a set of data that is close to uniformly distributed throughout a hypercube without requiring a dense grid of points that scales exponentially with the dimensionality. When sampling a Latin hypercube of $N$ points, each dimension is split into $N$ even segments, and the points are then placed within the resulting $N^d$ cells, where $d$ is the dimensionality, in a way that ensures that only a single cell in a set of $d$ rows is occupied by a point.

The Latin hypercube sampling does not, however, guarantee uniform sets of training data, given that having all points on a diagonal line also constitutes a Latin hypercube, but in practice, one will always get something very close to uniformity for a large amount of data points ($N>10^3$). A variant of Latin hypercube sampling called \textit{orthogonal sampling} furthermore ensures uniformity in the hypercube by splitting the cube into smaller segments~\cite{Tang93}. This will result in uniformity on large scales (similar to the size of the Latin hypercube), but points on smaller scales tend to look more randomly distributed. This is due to most implementations of Latin hypercube sampling placing points randomly within the $N^d$ cells~\cite{Iman80}. Orthogonal sampling is more complex than Latin hypercube sampling and the added guarantee is not worth the extra layer of complexity, since Latin hypercubes virtually always produces a set of points that are close to uniformly distributed on large scales.

Having a Latin hypercube as training data thus ensures that the resulting neural network will be equally precise in all parts of the parameter space, which is beneficial if one wants to remain completely agnostic with respect to cosmological models and data. A drawback when sampling from a hypercube, however, is that the density of points around the region of interest (when computing the likelihood function) becomes very small in higher dimensions. Most of the volume is in the corners of the hypercube in higher dimensions, and this leads to a very sparse sampling for all feasible numbers of points, $N$. This means that possible features in the best-fit region are not resolved very well. A neural network trained on a hypercube of points still gains information from the outermost points, but it usually requires many more epochs of training (Ref.~\cite{Bonici:2023xjk} reports 50,000 epochs for $10^4$ points) in order to extract all the required information to perform well in the region of interest. Hence, only very smooth likelihoods can be accurately represented by this sparse sampling. If the likelihood is highly non-Gaussian, the information from the sparsely sampled points might be insufficient for accurate emulation.

\subsection{Latin hypersphere sampling}\label{sec:latin_hypersphere}
\begin{figure}
	\centering
	\includegraphics[width=\textwidth]{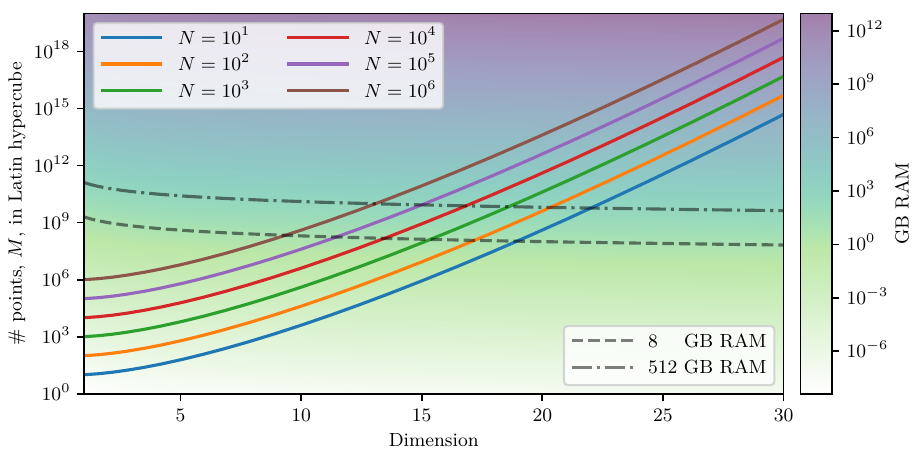}
	\caption{The figure shows how the ratio between the volumes of a hypercube and its inscribed hypersphere grows with higher dimensionality. The number of points, $M$, needed in a Latin hypercube in order to have $N$ points within the hypersphere is depicted as a function of dimensionality for various values of $N$. The background colours indicate the minimum required RAM in order to store a Latin hypercube of $M$ points of a certain dimension in memory (single precision), and two specific values of 8 GB and 512 GB have been highlighted by the dashed and dash-dotted lines, respectively.}
	\label{fig:rejection}
\end{figure}

This issue of the Latin hypercube can be circumvented by concentrating the sampling inside a hypersphere centred around the (approximate) best-fitting set of parameters. However, whereas the construction of the Latin hypercube is trivial, the procedure of sampling from a Latin hypersphere can be challenging in high dimensions. To illustrate, for generating a Latin hypersphere of $N$ points, one might naively think that a good solution is to generate a Latin hypercube of $M$ points, where $M = N\times r_d$ and $r_d$ is the ratio of the volume of a $d$-dimensional hypercube to that of an inscribed hypersphere \cite{Nygaard:2022wri},
\begin{equation}
	r_d = \frac{V_d^{\rm cube}}{V_d^{\rm sphere}} = \frac{2^d\,\Gamma(\frac{d}{2}+1)}{\pi^{\nicefrac{d}{2}}},
\end{equation}
and then reject all points with a Euclidean distance larger than 1 (radius of the hypersphere) to the centre. This is not ideal, however, since $r_d$ grows to very large values for higher dimensions as seen in figure~\ref{fig:rejection}. For example, if one needs to sample $10^4$ points in a 15-dimensional hypersphere, the required Latin hypercube would take up $\sim$50 GB of memory. In practice, this makes it very unfeasible to go beyond 10 dimensions, and outright impossible to go beyond 15 dimensions. 

However, since Latin hypercubes appear random on small scales, we might not need to enforce the Latin criterion on the sphere. If we can sample enough uniformly random points in the hypersphere, the density of points close to the best-fit region is still much greater than for any feasible Latin hypercube, and this set of points contains much more relevant information that can be extracted by a neural network in significantly fewer epochs during training. The question is then how to sample from a uniform hyperspherical distribution.

\subsection{Random uniform sampling from a hypersphere}

A way to uniformly sample from a hypersphere is to sample from another isotropic distribution and transform the points to a sphere afterwards~\cite{krauth06}. A standard multivariate normal distribution (i.e. a multivariate normal distribution with the identity covariance matrix $C=\mathds{1}$ and zero mean $\vec{\mu}=\vec{0}$) is one such isotropic distribution. Hence, if we sample $N$ points from a $d$-dimensional standard normal distribution and divide the coordinates of all points with their Euclidean distance from the centre, we obtain a sample of points uniformly distributed on the surface of a $d$-dimensional hypersphere with radius 1. We then need to distribute the points evenly throughout the hypersphere by multiplying the points by new radii. These new radii should be sampled from a non-uniform distribution in the interval $[0,1]$ in order to account for more points required in the outer parts compared to around the centre. Each hyperspherical shell needs to be weighted by the volume in that shell (scaling as $r^{d-1}$), which means that we need to sample from the distribution $\pi(r) = d\, r^{d-1}$, where the dimension constitutes a normalisation factor. The Probability Integral Transformation~\cite{CaseBerg:01} implies that the cumulative distribution function of $\pi(r)$ is uniformly distributed between 0 and 1, and we thus get
\begin{equation}
	\Pi(r) = \int_0^r  \pi(x)\,{\rm d}x = \int_0^r d\,x^{d-1}\,{\rm d}x = r^d \sim \mathcal{U}(0,1)\,.
\end{equation}
This means that we can just sample $\Pi(r)=r^d$ uniformly and then take the $d^{\rm th}$ root of the samples in order to get the distribution of radii. This is summarised in the pseudo-code depicted in algorithm~\ref{alg:hypersphere}.

\begin{pseudo}{Pseudo-code for random uniform sampling directly from a hypersphere. A similar algorithm is presented in Ref.~\cite{krauth06}.}{alg:hypersphere}
	\hspace{-1.4em}
	\begin{minipage}{0.52\textwidth}
		\vspace{1.7em}
		\begin{spacing}{1.2}
			\begin{algorithmic}
				\State $N$\tabto{2.5em}$=$\hspace{0.5em} number of points to sample
				\State $d$\tabto{2.5em}$=$\hspace{0.5em} dimensionality
				\State $S$\tabto{2.5em}$=$\hspace{0.5em} \texttt{Normal(num=N, dim=d)} 
				\State $R$\tabto{2.5em}$=$\hspace{0.5em}$\; \sqrt{{\rm sum}(S^2)}$ 
				\State $\Pi$\tabto{2.5em}$=$\hspace{0.5em} \texttt{RandomUniform([0, 1], num=N)}
				\State $R_{\rm new}$\tabto{2.5em}$=$\hspace{0.5em}$\; \Pi^{\nicefrac{1}{d}}$   
				\State $S$\tabto{2.5em}$=$\hspace{0.5em}$\; S \times R_{\rm new} / R$
			\end{algorithmic}
		\end{spacing}
	\end{minipage}
	&
	\hspace{0.07em}
	\begin{minipage}{0.45\textwidth}
		\vspace{4.4em}
		\begin{spacing}{1.2}
			\textit{draw points from a normal distribution\\
			compute  distances to the centre\\
			sample $\Pi(r)=r^d$ uniformly\\
			compute new radii\\
			rescale points to lie within the sphere}
		\end{spacing}
	\end{minipage}
\end{pseudo}

\subsection{Sampling near boundaries}\label{sec:boundaries}
In some cases we cannot sample from the entire hypersphere: If a parameter has a (physically motivated) hard boundary slicing the hypersphere, e.g. a non-negative particle mass, we cannot allow points outside of this boundary. In this case, the parameter space boundaries can be enforced through rejection sampling. One would then first sample from the hypersphere as if all of the parameter space is allowed, and then reject all points outside the boundaries. In practice, we implement this rejection sampling as a Python generator that maintains a cache of points generated using algorithm~\ref{alg:hypersphere}.

%This, of course, means that we end up with less points than we wanted, and to solve this, one can perform the sampling and rejection in a loop until the correct amount of points is obtained. This can, however, lead to a slower sampling if only a small fraction of the hypersphere is allowed.

%Alternatively, one can calculate the ratio, $r_A$ of the total volume of the hypersphere to the volume of the accepted piece and start with $N\times r_A$ points sampled in the unbounded hypersphere, where $N$ is the number of points we wish to have sampled in the end. This will, on average, yield an amount of accepted points closer to the desired amount than the naïve method. However, due to the randomness of the sampling, this will still not guarantee that we end up with $N$ points, so a combination of starting with more points and using a loop might be the best way to go\footnote{In our implementation, a generator with a large buffer size inside a while-loop is used for this purpose.}. 

Contrary to the rejection sampling described in section~\ref{sec:latin_hypersphere}, this is not expensive memory-wise since we can reject points on-the-fly. However, it might become computationally expensive if the boundaries of any parameter are such that only a thin slice of the hypersphere is allowed. In this case, the rejection rate would be close to 100\%. However, this situation could be easily remedied by sampling uniformly from this thin slice along the parameter in question (good approximation for thin slices) while still sampling from a hypersphere in the other parameters.

\subsection{Hyperellipsoid with correlations}
Finally, when sampling from a hypersphere, one will obviously need to scale the dimensions to fit the parameters (like one would scale a Latin hypercube), thus turning the hypersphere into a hyperellipsoid. This ellipsoid is uncorrelated in all parameters by construction, and in most cases this will be a very good sample, as we will see in section~\ref{sec:comparison}. We can, however, use additional information (if available) about correlations between parameters to significantly improve the performance by sampling along the known directions of correlation. In order to include correlations, we transform the sampled points from algorithm~\ref{alg:hypersphere} using the \textit{Cholesky transformation}~\cite{Press2007,Lewis:2013hha}. With the prior knowledge of the parameter correlations stored in the covariance matrix $C$, the lower triangular matrix $L$ that satisfies $C=LL^{\rm T}$ is determined from the Cholesky decomposition. From this, a transformed (correlated) point, $\tilde{p}$, is computed as the matrix multiplication $\tilde{p} = L p$, where $p$ is the uncorrelated point. This procedure, along with the rejection of points outside the parameter bounds described in section~\ref{sec:boundaries}, is summarised in algorithm~\ref{alg:correlated}. It is usually better to start with a known $\Lambda$CDM covariance matrix and then treat other parameters as uncorrelated if no information is available about their correlations than to not use any correlations at all. This will be explored in section~\ref{sec:comparison}.

\begin{pseudo}{Pseudo-code for random sampling in a correlated hyperellipsoid. This procedure uses a covariance matrix to transform points computed by algorithm~\ref{alg:hypersphere} to reflect correlations of the parameters.}{alg:correlated}
	\hspace{-1.4em}
	\begin{minipage}{0.53\textwidth}
		\vspace{1.7em}
		\begin{spacing}{1.2}
			\begin{algorithmic}
				\State $N$\tabto{3.6em}$=$\hspace{0.5em} number of points to sample
				\State $d$\tabto{3.6em}$=$\hspace{0.5em} dimensionality
				\State $C$\tabto{3.6em}$=$\hspace{0.5em} covariance matrix
				\State $M$\tabto{3.6em}$=$\hspace{0.5em} buffer size larger than $N$
				\State $L$\tabto{3.6em}$=$\hspace{0.5em} \texttt{CholeskyDecomposition(C)}
				\State $points$\tabto{3.6em}$=$\hspace{0.5em} \{ \}
				\While{\texttt{Size(}$points$\texttt{)} < $N$}
					\State $P$\tabto{3.6em}$=$\hspace{0.5em} \texttt{Hypersphere(M,d)} 
					\State $P$\tabto{3.6em}$=$\hspace{0.5em} \texttt{MatrixMultiplication(L,P)} 
					\State $P_{\rm ac}$\tabto{3.6em}$=$\hspace{0.5em} \{$p$ within boundaries for $p$ in $P$\}
					\State append $P_{\rm ac}$ to $points$
				\EndWhile
				\State $points$\tabto{3.6em}$=$\hspace{0.5em} select $N$ elements of $points$
			\end{algorithmic}
		\end{spacing}
	\end{minipage}
	&
	\hspace{0.07em}
	\begin{minipage}{0.44\textwidth}
		\vspace{5.92em}
		\begin{spacing}{1.2}
			\textit{buffer to handle rejection sampling\\
			compute lower triangular matrix\\
			initialise set of points\\
			\\
			get $M$ points from algorithm~\ref{alg:hypersphere}\\
			transform points\\
			accept only points within boundaries}
		\end{spacing}
		\begin{spacing}{1.18}
			\textit{\\
			keep only $N$ points}
		\end{spacing}
	\end{minipage}
\end{pseudo}

\section{Comparisons using {\bfseries\scshape connect}}\label{sec:comparison}

\begin{table}[t]
	\begin{center}
		\begin{tabular}{l|ll}
			               & \textbf{Lower boundary} & \textbf{Upper boundary} \\ \hline
			$\omega_{\rm b}$     & 0.014          & 0.039          \\
			$\omega_{\rm cdm}$     & $10^{-11}$     & 0.25           \\
			$H_0$                & 30             & 120            \\
			$\log(10^{10}A_s)$   & 1              & 5              \\
			$n_s$                & 0.7            & 1.3            \\
			$\tau_{\rm reio}$    & 0.01           & 0.4            \\
			$f_{\rm EDE}$        & $10^{-11}$     & 0.3            \\
			$\log_{10}(z_c)$     & 3              & 4.3            \\
			$\theta_i^{\rm scf}$ & 0.1            & 3.1            \\
			$m_{\rm ncdm}$       & 0.02           & 10             \\
			$N_{\rm ur}$         & 0              & 6             
		\end{tabular}
	\end{center}
	\caption{Lower and upper boundaries of the cosmological parameters. These are used for the initial Latin hypercubes and hyperspheres, and they are also enforced during the MCMC samplings. The first 6 parameters constitute the $\Lambda$CDM model, while all 11 parameters constitute the EDE+$M_\nu$+$N_\mathrm{ur}$ model.}
	\label{tab:boundaries}
\end{table}

In order to show the benefits of hypersphere sampling, we have tested this against the more conventional approach of using Latin hypercubes. We use the \connect{} framework\footnote{Publicly available at \url{https://github.com/AarhusCosmology/connect_public}.}~\cite{Nygaard:2022wri} to sample training data and train our neural networks. The sampling of training data is done using the iterative approach of \connect{}, where an initial neural network is trained on sparse uniformly distributed points (Latin hypercubes up until now) and then used to sample new points using MCMC. This continues until a set of training data points, representative of the likelihood, is built. The new implementation allows for a different sampling of the points for the initial model than the regular Latin hypercube sampling. Instead, one can now use a uniformly sampled hypersphere, as presented in section~\ref{sec:sampling}, as a starting point for the iterative process. In this section we will explore how well the initial neural networks (using hyperspheres or Latin hypercubes) as well as the networks from the final iteration of the iterative sampling emulate the output from \class{}. We present results from two different cosmological models, i.e. the standard 6-parameter $\Lambda$CDM model and a 5-parameter extension with Early Dark Energy (EDE)~\cite{Karwal:2016vyq,Poulin:2018cxd,Poulin:2023lkg}, additional ultra-relativistic degrees of freedom $N_\mathrm{ur}$ and massive neutrinos $M_\nu$ to really challenge both the \connect{} framework as well as the two methods of initial sampling. For each cosmological model, we compare the following initial configurations:
\begin{itemize}
	\item \textbf{HS (correlated):} Hypersphere with 1,000 points with correlations from a converged MCMC run.
	\item \textbf{HS (uncorrelated):} Hypersphere with 1,000 points with no correlations.
	\item \textbf{HS ($\Lambda$CDM correlated):} Hypersphere with 1,000 points with $\Lambda$CDM correlations and no correlations for the extended parameters (only applies to the EDE+$M_\nu$+$N_\mathrm{ur}$ runs).
	\item \textbf{LHC (Small):} Latin hypercube with 1,000 points.
	\item \textbf{LHC (Medium):} Latin hypercube with 10,000 points.
	\item \textbf{LHC (Large):} Latin hypercube with 100,000 points.
\end{itemize}
The boundaries in the parameter space can be seen in table~\ref{tab:boundaries}. All networks have been trained for 500 epochs with a batchsize of 256, and otherwise the same hyperparameters as in Ref~\cite{Nygaard:2022wri}. The training data from the first iterations (using the initial neural networks) has been discarded for all Latin hypercubes (standard setting in \connect{}), since it is usually far from the best-fit region and therefore contaminates our total set of training data, while data from all iterations has been kept for the hyperspheres. For the subsequent MCMC runs using the neural networks, we have employed a data set consisting of:
\begin{itemize}
	\item Planck 2018 high-$\ell$ TTTEEE, low-$\ell$ TT+EE, and lensing~\cite{Planck:2018vyg,Planck:2019nip}.
	\item Baryon Acoustic Oscillations (BAO) measurements from BOSS DR12~\cite{boss2016}, the main galaxy sample of BOSS DR7~\cite{ross2014} and 6dFGS~\cite{Beutler:2011hx}.
	\item Pantheon supernova data~\cite{Pan-STARRS1:2017jku}.
\end{itemize}
The training data is gathered using the marginalised Planck 2018 high-$\ell$ TTTEEE Lite likelihood due to its rapid evaluation time along with low-$\ell$ TT+EE. This data set is less constraining and always produces adequate training data~\cite{Nygaard:2022wri} when the final data set includes the full Planck likelihood. For each MCMC, we have run 6 chains using \montepython{}~\cite{Audren:2012wb,Brinckmann:2018cvx}, considering the runs to be converged when the Gelman-Rubin statistic fulfils $R-1<0.01$. Some of the MCMC runs for the EDE+$M_\nu$+$N_\mathrm{ur}$ model using the initial neural networks trained on either hyperspheres or Latin hypercubes have difficulties converging, and was stopped when the number of accepted points were similar to the converged runs.

\subsection{$\Lambda$CDM}
\begin{figure}
	\begin{center}
		\includegraphics[width=\textwidth]{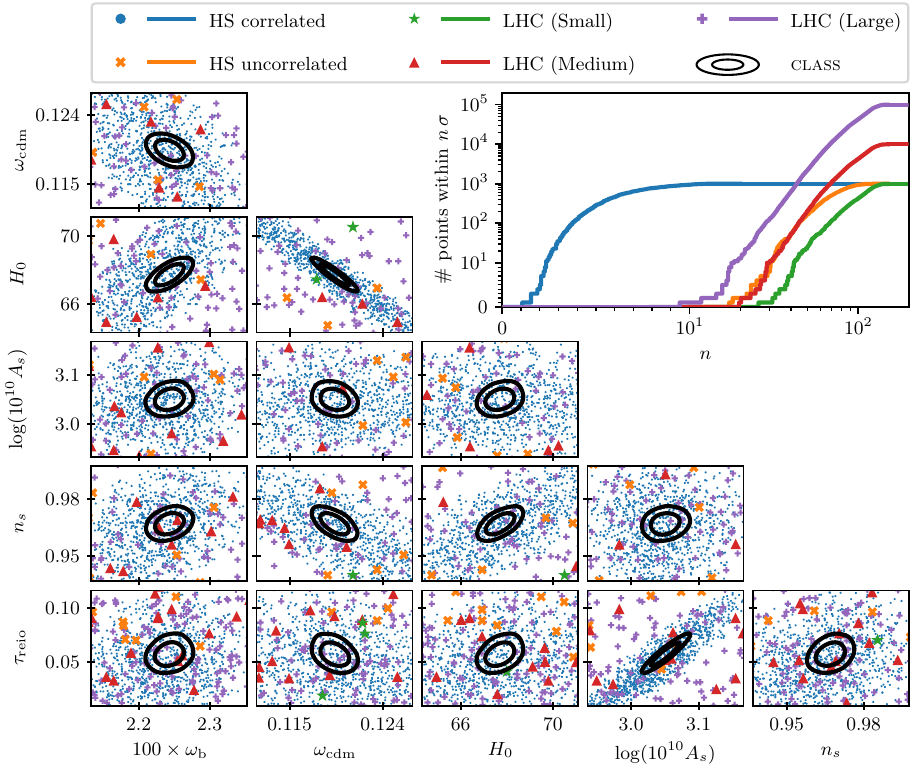}
		\caption{The triangle plot shows the point distributions of the various initial configurations in the $\Lambda$CDM model within 50 standard deviations of the best-fit point in all parameters along with the 2D posteriors when using \class{}. The top-right panel shows the number of points within $n$ standard deviations of the best-fit point in all parameters as a function of $n$ for the different configurations.}
		\label{fig:points}
	\end{center}
\end{figure}

The $\Lambda$CDM model has the parameter vector $\vec{\Theta} = \{\omega_{\rm b},\,\omega_{\rm cdm},\, H_0,\, \log(10^{10}A_s),\, n_s,\, \tau_{\rm reio}\}$ along with extra relativistic relics that fix $N_{\rm eff} = 3.046$. This is often quite easy to sample because of its likelihood surface being almost perfectly Gaussian. Figure~\ref{fig:points} highlights the differences in sampling density around the best-fit region between the various initial configurations. The triangle plot only includes points that are within 50 standard deviations (determined by an MCMC with \class{}) of the best-fit point in all parameters. It is evident that the correlated hypersphere has a much higher point density around the contours of the \class{} posterior, with the largest Latin hypercube with 100,000 points being second. It is also worth noticing that the uncorrelated hypersphere and the medium Latin hypercube with 10,000 points have roughly similar point densities. This is also supported by the top-right panel, which shows the number of points within $n$ standard deviations of the best-fit point in all parameters as a function of $n$ for the different configurations, where the uncorrelated hypersphere and medium Latin hypercube follow each other up until around $n=50$. The limit of 50 standard deviations in the triangle plot was chosen such that the Latin hypercubes would have around half of their points included. If we had chosen too include all points, the view of the best-fit region would have been obscured by the large amount of points from the Latin hypercubes that are close to the best-fit in the 2 parameters of each 2D plane, but very far away in other parameters. The top-right panel also clearly shows how much closer to the best-fit point the points of the correlated hypersphere are. One would need a Latin hypercube of many orders of magnitude more points to achieve the same density.

\begin{figure}
	\centering
	\includegraphics[width=\textwidth]{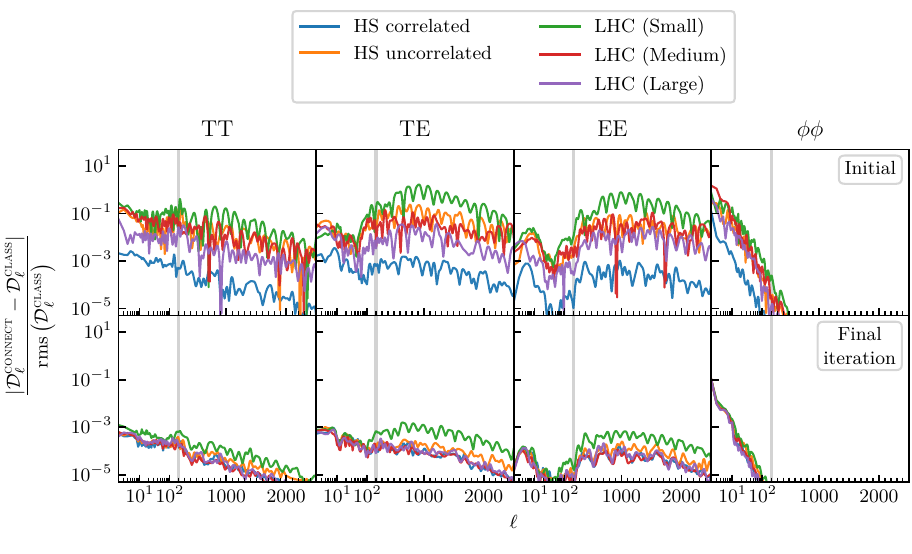}
	\caption{Errors of the neural networks emulating the $\Lambda$CDM model when emulating the CMB spectra of representative test data. The top panels show the errors of the initial models before the iterative process and the bottom panels show the errors of the networks from the final iterations of their respective runs. The test data is taken from a converged $\Lambda$CDM MCMC and is therefore the error around the best-fit and not an indicator of the training error. The lines correspond to a 95.45\% confidence level where 95.45\% of the computed points have errors beneath the lines.}
	\label{fig:lcdm_error}
\end{figure}

\begin{figure}[h]
	\centering
	\includegraphics[width=\textwidth]{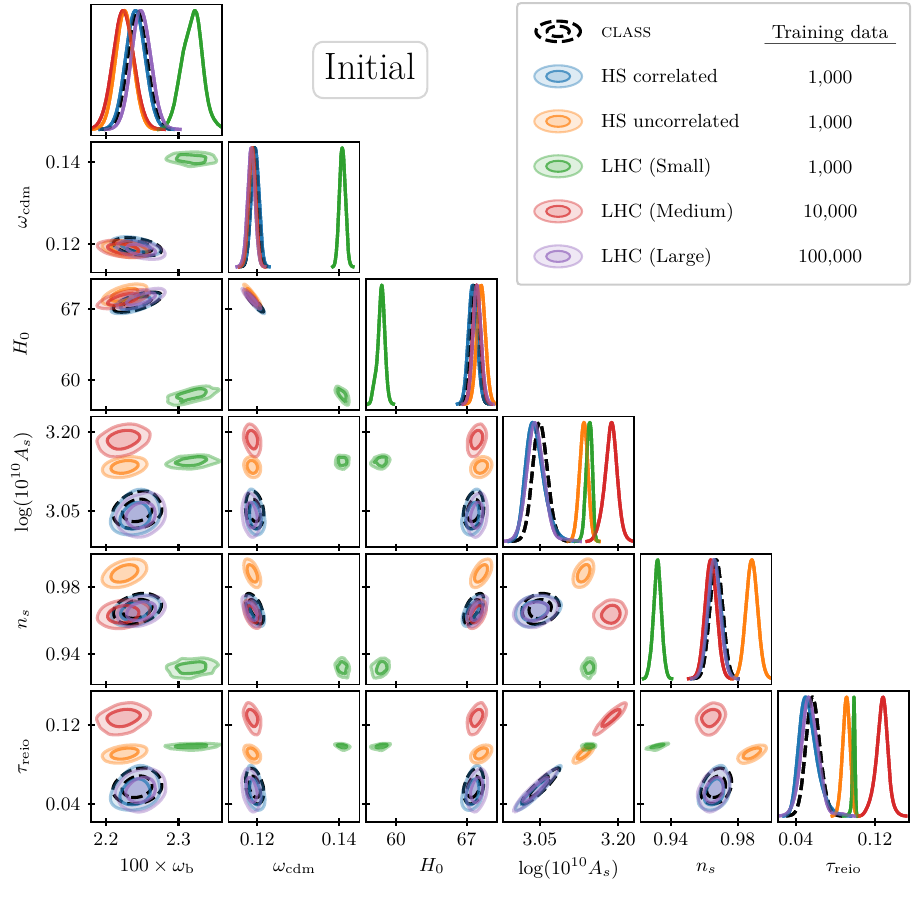}
	\caption{Posteriors from the MCMC runs using the initial neural networks emulating the $\Lambda$CDM model. A similar MCMC run using \class{} is also shown in black dashed lines as a reference. The table in the legend gives the number of training data points the respective networks used for the MCMCs were trained on.}
	\label{fig:lcdm_mcmc_ini}
\end{figure}

\begin{figure}[h]
	\centering
	\includegraphics[width=\textwidth]{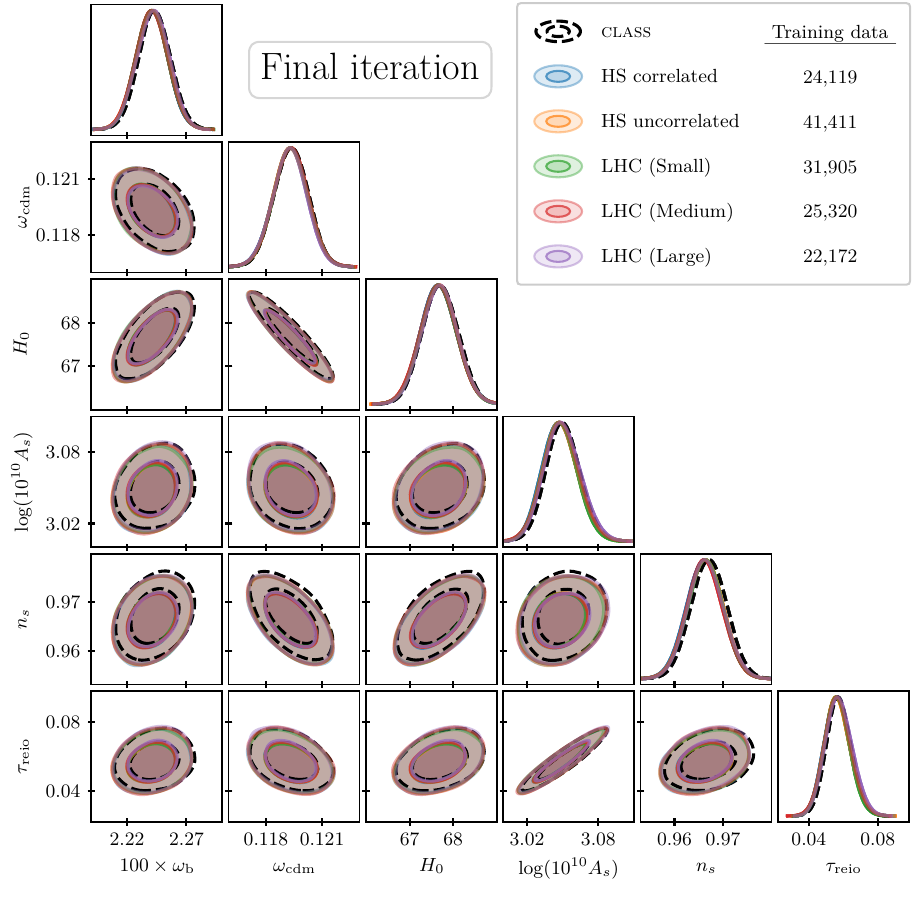}
	\caption{Posteriors from the MCMC runs using the neural networks emulating the $\Lambda$CDM model from the final iterations. A similar MCMC run using \class{} is also shown in black dashed lines as a reference. The table in the legend gives the number of training data points the respective networks used for the MCMCs were trained on.}
	\label{fig:lcdm_mcmc_final}
\end{figure}

Figure~\ref{fig:lcdm_error} shows, for each of the TT, TE, EE and $\phi \phi$ spectra, the difference in the $\mathcal{D}_\ell \equiv \ell (\ell + 1) C_\ell / 2\pi$ coefficients between \connect{} and \class{} relative to their root-mean-square values in \class{}. The top panel shows the 95.45\% percentile of the error based on only the initial configurations while the bottom panel shows the same for the final iterations. It is apparent that a neural network trained on a correlated hypersphere outperforms all of the other initial networks in terms of preciseness around the best-fit region, since it has the most representable training data. In fact, in order to be as precise a hypersphere with only 1,000 points, a Latin hypercube would need more than 2 orders of magnitude more points (when restricting the training process to 500 epochs). Even if we lose the information of any correlations, the uncorrelated hypersphere seems to be just as good as a Latin hypercube with 10 times the amount of points, as also suggested by figure~\ref{fig:points}. 

All of the runs, however, converge on good training data as can be seen in the lower panels of figure~\ref{fig:lcdm_error}, where the precisions of the networks from all of the final iterations are shown. The only one that stands out as slightly less precise after the final iteration is the one with an initial Latin hypercube of 1,000 points. This is most likely due to it having more training data away from the best-fit region. The initial network of this run is also worse than the rest, and it probably takes more iterations to locate the best-fit region than just the first one that is discarded afterwards. This leaves us with a lot of training data far from the region of interest, thus making the network less precise in order to accommodate the additional training data. 

Figures~\ref{fig:lcdm_mcmc_ini} and \ref{fig:lcdm_mcmc_final} show the posteriors obtained from emulators trained on the initial and final configurations, respectively, along with posteriors obtained using \class{}. We can again see that the initial networks trained on a correlated hypersphere of 1,000 points and a Latin hypercube of 100,000 points perform similarly, while the initial network trained on an uncorrelated hypersphere of 1,000 points performs similarly to one trained on a Latin hypercube of 10,000 points. The small Latin hypercube of only 1,000 points is too sparse to give a good starting point, since its contours are far away from the rest. This is, however, not a problem when using the iterative approach, as is apparent from figure~\ref{fig:lcdm_mcmc_final} where the posteriors from the final configuration emulators are seen to be nearly identical; instead, the downside of a poor initial configuration is that it requires more iterations and training data, reducing the total gain in computational efficiency from the emulation. 

\begin{table}[t]
	\begin{center}
		\begin{tabular}{l|ll}
			& \textbf{Iterations} & {\bfseries\scshape class} \textbf{evaluations} \\ \hline
			\textbf{Correlated hypersphere}   & 3                    & 25,119           \\
			\textbf{Uncorrelated hypersphere} & 3                       & 42,411            \\
			\textbf{Latin hypercube (Small)}  & 5                     & 37,905           \\
			\textbf{Latin hypercube (Medium)} & 4                 & 40,320           \\
			\textbf{Latin hypercube (Large)}  & 3                & 127,172         
		\end{tabular}
	\end{center}
	\caption{Number of iterations and amount of \class{} evaluations in each $\Lambda$CDM \connect{} run. The initial data from hyperspheres and Latin hypercubes is always discarded and furthermore the first iteration of 5,000 points is discarded for all Latin hypercubes.}
	\label{tab:lcdm_points}
\end{table}
Table~\ref{tab:lcdm_points} shows the final number of iterations in each \connect{} training procedure along with how many \class{} evaluations have been used. The initial hypersphere and Latin hypercube data is always discarded and the first iteration consisting of 5,000 points is discarded for all Latin hypercubes. We can see that the run with the small Latin hypercube indeed takes more iterations, but the amount of \class{} evaluations is similar to the runs with the uncorrelated hypersphere and the medium Latin hypercube. Although the number \class{} evaluations is a good way to measure roughly how much CPU time is spent (since it is the slowest part of the sampling), there is an overhead from the MCMCs and training of each iteration that become more significant with several iterations. With a close-to-Gaussian likelihood like $\Lambda$CDM (or simple extensions), the configuration does not matter much for the final result when using relatively low precision settings (e.g. a low number of epochs), but one can significantly speed up the process and reduce the computational cost by using a hypersphere instead of a Latin hypercube.

\subsection{EDE+$M_\nu$+$N_\mathrm{ur}$}
This large extension model with a parameter vector of  $\vec{\Theta} = \{\omega_{\rm b},\,\omega_{\rm cdm},\, H_0,\, \log(10^{10}A_s),\, n_s,\, \tau_{\rm reio},\, f_{\rm EDE},\, \log_{10}(z_c),\, \theta_i^{\rm scf},\, m_{\rm ncdm},\, N_{\rm ur}\}$ consists of the usual $\Lambda$CDM parameters, two massless neutrinos, a single neutrino with mass $m_{\rm ncdm}$, additional ultra-relativistic degrees of freedom contributing a value $N_\mathrm{ur}$ to the amount of relativistic degrees of freedom in the early Universe, and finally, an early dark energy (EDE) model~\cite{Poulin:2023lkg}. The particular EDE model used here is the original axion-like model based on~\cite{Karwal:2016vyq,Poulin:2018cxd}, which involves an axion-like scalar field that is frozen at its initial field value $\theta^\mathrm{scf}_i$ due to Hubble friction, until a redshift $z_c$, at which it rolls to the bottom of its potential, acting effectively as a fastly decaying fluid. Since it acts as a vacuum energy initially, its maximum fractional contribution to the energy budget, $f_\mathrm{EDE}$, is realised at the decay time $z_c$. In the following, we use the implementation of the EDE model presented in~\cite{Hill:2020osr}\footnote{Publicly available at \url{https://github.com/mwt5345/class_ede}.}. This large, combined model was primarily chosen to showcase the potential of using hyperspheres instead of Latin hypercubes, and it also serves as a good test of the \connect{} framework. 

\begin{figure}
	\centering
	\includegraphics[width=\textwidth]{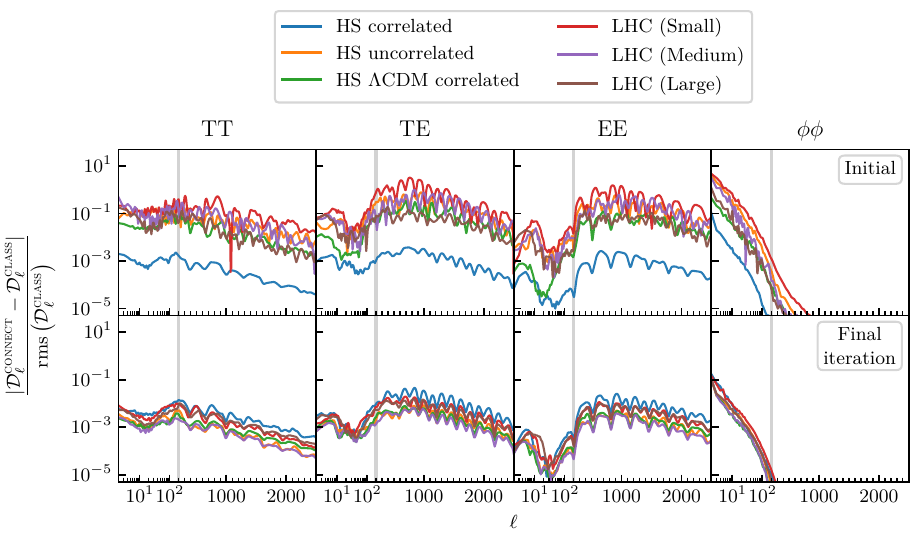}
	\caption{Errors of the neural networks emulating the EDE+$M_\nu$+$N_\mathrm{ur}$ model when emulating the CMB spectra of representative test data. The top panels show the errors of the initial models before the iterative process and the bottom panels show the errors of the networks from the final iterations of their respective runs. The test data is taken from a converged EDE+$M_\nu$+$N_\mathrm{ur}$ MCMC and is therefore the error around the best-fit and not an indicator of the training error. The lines correspond to a 95.45\% confidence level where 95.45\% of the computed points have errors beneath the lines.}
	\label{fig:ede_error}
\end{figure}

The precision settings are the same as before with 500 epochs during training of the networks even though the MCMCs with this model take much more time to converge due to it being far from Gaussian in the extended parameters. Figure~\ref{fig:ede_error} shows the 95.45\% percentiles of the errors in the CMB coefficients emulated by the initial configuration networks (top panel) and the networks from the final iterations (bottom panel) from \connect{}, compared to the values obtained directly from \class{}. It is clear from the figure that the initial neural network trained on a correlated hypersphere using the actual correlations of the model is much more precise than the rest of the initial networks within the best-fit region. Again we see a similar performance of the uncorrelated hypersphere and the medium Latin hypercube with 10,000 points, but in addition we also see a similar performance between the hypersphere with only $\Lambda$CDM correlations (and no correlations in the extended parameters) and the large Latin hypercube with 100,000 points. 
%After the final iterations, the correlated hypersphere and the large Latin hypercube seem to be less precise around the best-fit region than the rest despite having a better starting point before the iterations. This could, however, be due to their training data capturing more of the wider correlations of the model and thus having slightly less precision around the best-fit. This could of course be remedied by increasing the number of epochs during training of the last neural network. 

\begin{figure}[t]
	\centering
	\includegraphics[width=\textwidth]{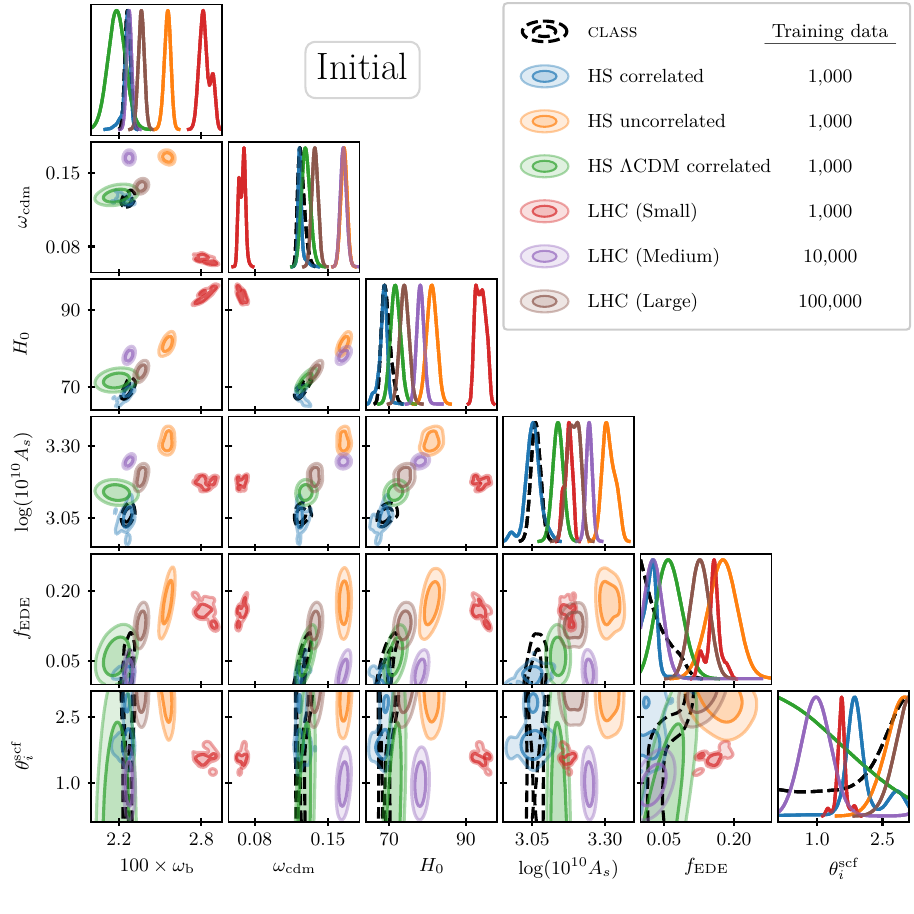}
	\caption{Posteriors from the MCMC runs using the initial neural networks emulating the EDE+$M_\nu$+$N_\mathrm{ur}$ model. A similar MCMC run using \class{} is also shown in black dashed lines as a reference. The table in the legend gives the number of training data points the respective networks used for the MCMCs were trained on.}
	\label{fig:ede_mcmc_ini}
\end{figure}

Figure~\ref{fig:ede_mcmc_ini} shows the posteriors from the initial neural networks of all runs. A mix between $\Lambda$CDM parameters and extended parameters have been chosen that best depicts how far some of the posteriors are from the best-fit region, since many of the extended parameters have significant posteriors all throughout their prior bounds. We clearly see that the correlated hypersphere has the most overlap with \class{}, although 1,000 points and 500 epochs is too small to correctly emulate the model. The hypersphere with only $\Lambda$CDM correlations seems to be the one with the second most overlap, even though some correlations differ significantly from the $\Lambda$CDM model. This suggests that it is reasonable to use $\Lambda$CDM correlations if true correlations are not known beforehand, even though the model is significantly different. The small Latin hypercube once again performs worse than the others, and it is much too sparse to discover any correlations in the model which is apparent from the ($H_0$,$\log(10^{10}A_s)$)--contour where all posteriors fall on the same line except that of the small Latin hypercube.

\begin{figure}[h]
	\centering
	\includegraphics[width=\textwidth]{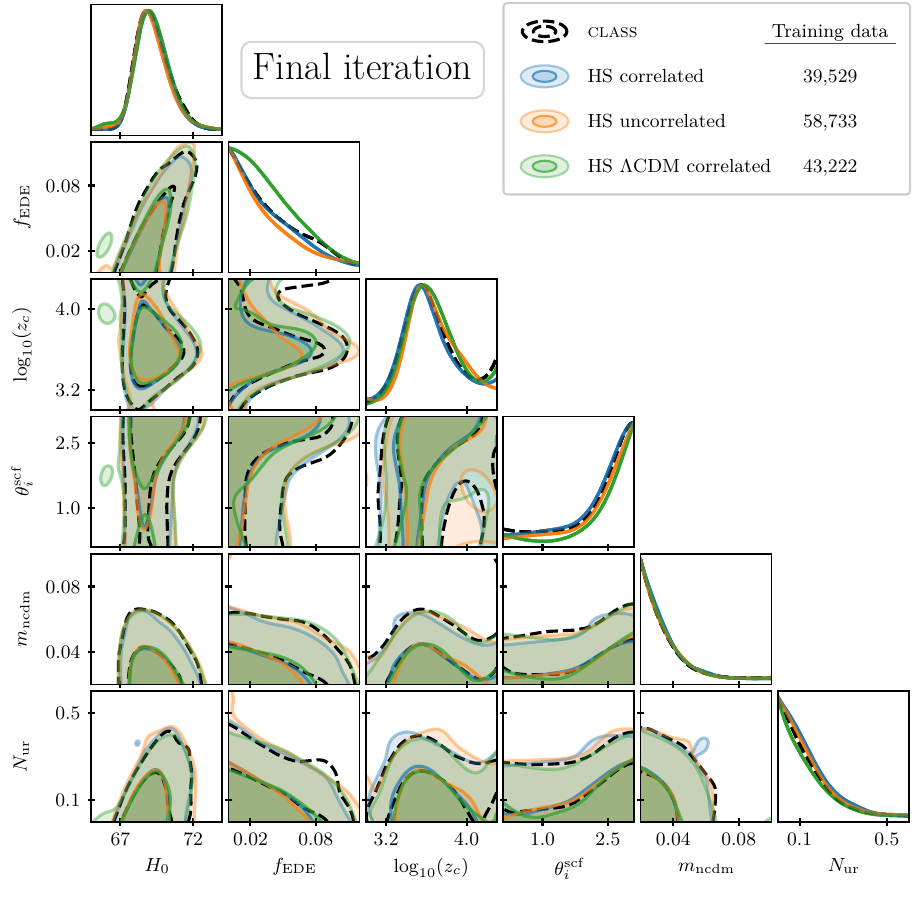}
	\caption{Posteriors from the MCMC runs using the neural networks emulating the EDE+$M_\nu$+$N_\mathrm{ur}$ model from the final iterations of the hypersphere runs. A similar MCMC run using \class{} is also shown in black dashed lines as a reference. The table in the legend gives the number of training data points the respective networks used for the MCMCs were trained on. }
	\label{fig:ede_mcmc_final_hs}
\end{figure}
\begin{figure}[h]
	\centering
	\includegraphics[width=\textwidth]{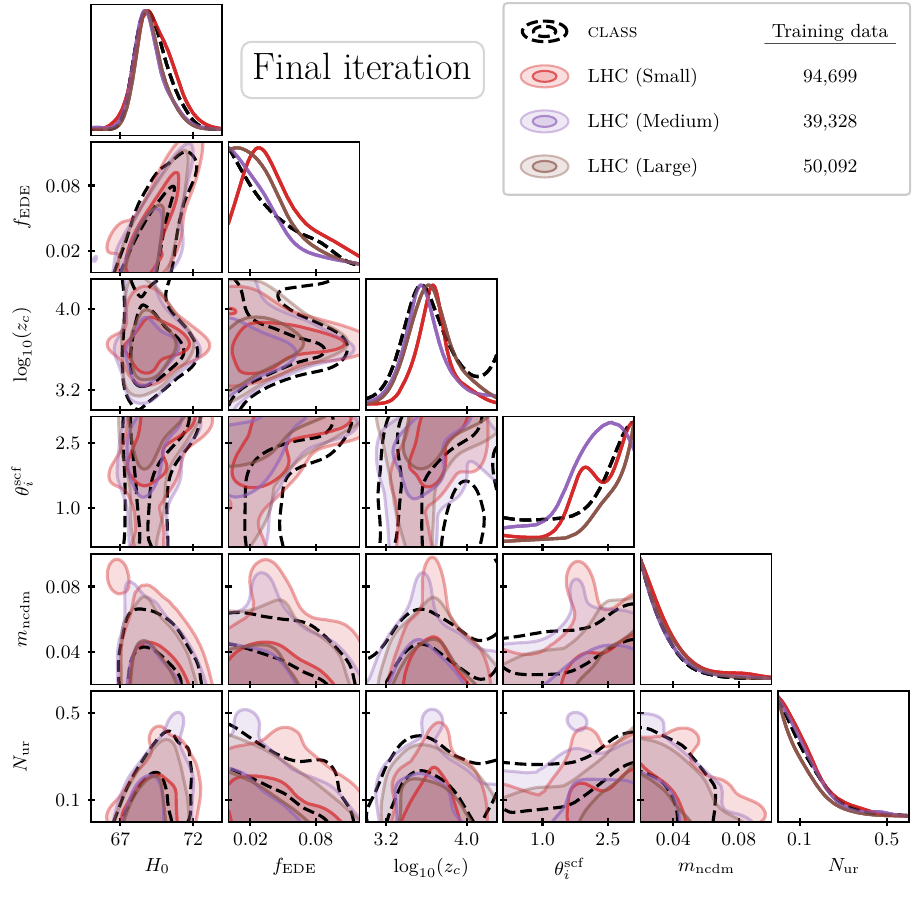}
	\caption{Posteriors from the MCMC runs using the neural networks emulating the EDE+$M_\nu$+$N_\mathrm{ur}$ model from the final iterations of the Latin hypercube runs. A similar MCMC run using \class{} is also shown in black dashed lines as a reference. The table in the legend gives the number of training data points the respective networks used for the MCMCs were trained on.}
	\label{fig:ede_mcmc_final_lhc}
\end{figure}

The figures~\ref{fig:ede_mcmc_final_hs} and~\ref{fig:ede_mcmc_final_lhc} show the posteriors when sampling with the neural networks from the final iterations for hyperspheres and the Latin hypercubes, respectively. Only the extended parameters along with $H_0$ are shown, since all other posteriors are close to Gaussian and agree very well with \class{} for all runs. It is immediately clear that the posteriors calculated using the final iterations of the hyperspheres are much closer to the \class{} results than those calculated using the Latin hypercubes. It seems that using a Latin hypercube as initial training data (even with 100,000 points) for a complicated cosmological model does not produce the correct result with the default settings of \connect{}. Training for more epochs and collecting more data from each iteration will of course solve this, but this would also increase the computational costs significantly. The correlated hyperspheres (true correlations and $\Lambda$CDM correlations) are slightly more accurate than the one with no correlations, and this again suggests that slightly wrong correlations might be better than no correlations at all. 

Indeed, the two correlated hyperspheres in figure~\ref{fig:ede_mcmc_final_hs} actually seem to agree quite well with \class{} and the subtle differences could be explained by the \class{} MCMC being slightly less converged. The curious "hole" in the $(\log_{10}{z_c},\theta_i^{\rm scf})$--contour is also reproduced by the final iterations of the correlated hyperspheres and to a lesser extend also the uncorrelated hypersphere. This is not lack of convergence, since we expect this feature to be present (see e.g.\ \cite{Hill:2020osr,Poulin:2023lkg,Hill:2021yec,McDonough:2023qcu}).

Table~\ref{tab:ede_points} shows the number of iterations and total amount of \class{} evaluations from all these \connect{} runs, and it apparent that the runs with the small and large Latin hypercubes took significantly more iterations. This typically indicates that the iterations "jump" around the parameter space and has difficulties homing in on the best-fit region. 

\begin{table}[]
	\begin{center}
		\begin{tabular}{l|ll}
			& \textbf{Iterations} & {\bfseries\scshape class} \textbf{evaluations} \\ \hline
			\textbf{Correlated hypersphere}     					& 5                    & 40,529           \\
			\textbf{Uncorrelated hypersphere}					& 4                    & 59,733            \\
			$\bm{\Lambda}$\textbf{CDM correlated hypersphere}	& 4                    & 44,222            \\
			\textbf{Latin hypercube (Small)}  					& 8                    & 100,699           \\
			\textbf{Latin hypercube (Medium)} 					& 5           		& 54,328           \\
			\textbf{Latin hypercube (Large)} 					& 7			& 155,092         
		\end{tabular}
	\end{center}
	\caption{Number of iterations and amount of \class{} evaluations in each EDE+$M_\nu$+$N_\mathrm{ur}$ \connect{} run. The initial data from hyperspheres and Latin hypercubes is always discarded and furthermore the first iteration of 5,000 points is discarded for all Latin hypercubes.}
	\label{tab:ede_points}
\end{table}

\begin{figure}
	\begin{center}
		\includegraphics[width=\textwidth]{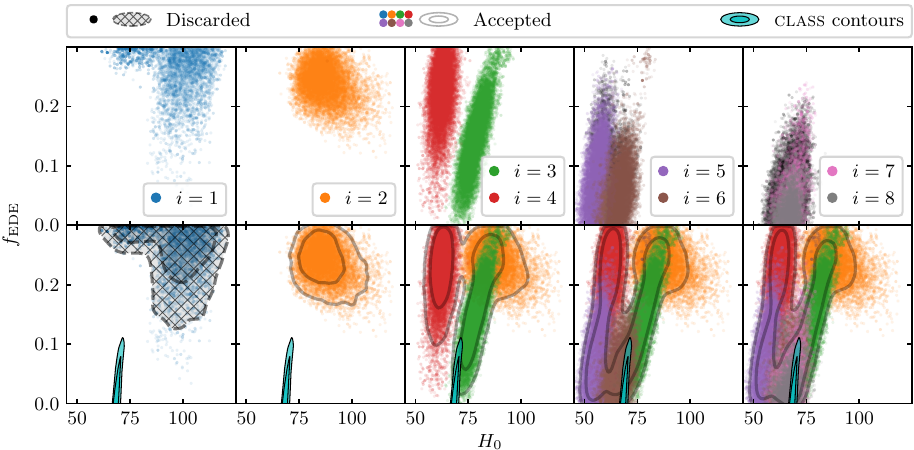}
		\caption{A depiction of how the training data was sampled iteratively in the \connect{} run with a small initial Latin hypercube of 1,000 points in the $(H_0,f_{\rm EDE})$--plane. The black points and the hatched contours represent discarded data. The \class{} contours have been included for reference.}
		\label{fig:iterations}
	\end{center}
\end{figure}

Figure~\ref{fig:iterations} illustrates the iterative training data sampling of \connect{} that has the small Latin hypercube ($10^3$ points) as initial configuration, in the ($H_0$,$f_\mathrm{EDE}$)--plane. It is seen that the training data gathered by the small Latin hypercube run has difficulties converging, resulting in training data from very different regions of the parameter space with little to no overlap in the first few iterations. This means that we end up with a contaminated set of training data where the neural network attempts to fit the large contamination during training at the cost of precision around the best-fit region (represented by contours from an MCMC using \class{} in the figure). Only in the final iterations does it seem to represent the best-fit region well. Usually, the iterations would overlap more, which filters away more points, so fewer \class{} evaluations are needed, but in this case, nearly all points are kept and points are only filtered out near the final iterations (aside from the first iteration which is completely discarded by default). There are fortunately simple solutions to accommodate this problem:

\begin{itemize}
	\item using higher precision settings, e.g. training for many more epochs,
	\item throwing away training data from more iterations than the first, even though it can be difficult to know in advance,
	\item restarting the \connect{} run with the last model from the previous run as the initial neural network.
\end{itemize}
These solutions will all be able to solve the problem, but at much greater computational cost. Training for many more epochs can significantly slow down the iterative process, throwing away too many \class{} evaluations is wasteful and should be avoided, and restarting the run means that all previous training data is discarded, which is also wasteful. The least wasteful approach with the lowest computational cost will therefore be to use a hypersphere as initial guess instead (with correlations if available) and perhaps use more than 1,000 points for complicated cosmological models with a high dimensionality like this one.

\section{Conclusion and outlook}\label{sec:conclusion}

We have tested the performance of networks trained on hyperspheres and Latin hypercubes as well as how these impact the performance of the iterative approach of the \connect{} emulation framework. It is apparent that the neural networks trained on hyperspheres greatly outperform the networks trained on Latin hypercubes of similar size and with the same hyperparameters. Although weak knowledge about the shape and location of the posterior is required to initialise the hypersphere, this is not much different from the knowledge needed to initialise a Latin hypercube. We find that even using an uncorrelated hypersphere instead of a Latin hypercube cuts the amount of training data required for the same performances down by an order of magnitude, increasing to several orders of magnitude if correlations are included. Even in the case of a high-dimensional and highly non-Gaussian cosmological model like the EDE+$M_\nu$+$N_\mathrm{ur}$ model, the correlated hypersphere with only 1,000 points trained for just 500 epochs was able to capture a lot of the behaviour of the observables under variations of the model parameters. These are quite low precision settings for the initial model, since its job in \connect{}'s iterative sampling is only to approximately locate the best-fit region, so if one wished to train a network just on a hypersphere without the iterative approach, an order of magnitude more points and epochs, should be sufficient in nearly all cases. No matter the initial configuration presented in this paper, using a final \connect{} network for MCMC runs is computationally much cheaper than using \class{} directly in the MCMC, including sampling of training data, training the network, and the MCMC itself. This is especially true for elaborate cosmological models with difficulties converging.

As shown by other emulators~\cite{SpurioMancini:2021ppk,Bonici:2023xjk}, Latin hypercubes are most certainly capable of good precision, but they require orders of magnitude more epochs, and for beyond-$\Lambda$CDM models also orders of magnitude more training data. If the aim of an emulator is to emulate a range in all parameters equally well, the Latin hypercube is still a good option, but if the aim is to use an emulator to compute likelihoods, the most effective option is to have the distribution of training data resemble the likelihood function. This is accomplished by the iterative approach of \connect{}, but one can get very close to the same effectiveness with a targeted uniform sampling representing the best-fit region, i.e. a correlated hypersphere. Especially in cases where the underlying cosmological code is prohibitively expensive, e.g. $N$-body codes, we conjecture that the performance would be greatly improved by switching to hypersphere sampling instead of Latin hypercube sampling. Indeed, emulators of $N$-body codes often only use very few training data points due to them being very slow to evaluate, but with a hypersphere, those few points would be much better distributed in order to capture the behaviour where the likelihood is significant. \\

\noindent {\bf Reproducibility.}
We have used the publicly available \connect{} framework available at \url{https://github.com/AarhusCosmology/connect_public} to create training data and train neural networks. The framework has been extended with the new way of sampling initial training data using hyperspheres. Explanatory parameter files have been included in the repository in order to easily use the framework and reproduce results from this paper.

\section*{Acknowledgements}
We acknowledge computing resources from the Centre for Scientific Computing Aarhus (CSCAA). A.N., E.B.H., and T.T. were supported by a research grant (29337) from VILLUM FONDEN. We would like to thank Alessio Spurio Mancini and Sven Günther for their feedback on the draft of this paper.

%%%%%%%%%%%%%%%%%%%%%%%%%%%%%%%%%%%%%%%%%%%%%%%%%%%%%%

%%%%%%%%%%%%
\bibliographystyle{utcaps}
%\nocite{*}
\bibliography{hypersphere2024}

\end{document}